\documentclass[fleqn,10pt]{wlscirep}
\usepackage[utf8]{inputenc}
\usepackage[T1]{fontenc}
\usepackage{multirow}
\usepackage{graphicx}
\usepackage{makecell}
\usepackage{array}
\title{Street access, Informality and Development: A block level analysis across all of sub-Saharan Africa}

\author[1,2,3,*]{Lu\'is M. A Bettencourt}
\author[2,*]{Nicholas Marchio}
\affil[1]{University of Chicago, Department of Ecology and Evolution, Chicago IL, 60637, USA}
\affil[2]{University of Chicago, Mansueto Institute for Urban Innovation, Chicago IL, 60637, USA}
\affil[3]{Santa Fe Institute, Santa Fe NM, 87501, USA}

\affil[*]{corresponding author(s): Lu\'is Bettencourt (bettencourt@uchicago.edu), Nicholas Marchio (nmarchio@uchicago.edu)}


\begin{abstract}
Sustainable development is an imperative worldwide but metrics and data on poverty and quality of life have remained too coarse and abstract to characterize challenges adequately and guide practical progress.  Nowhere is this challenge greater than in Africa where we still know relatively little about the systematic spatial details and scope of development. Here, we leverage a complete, high-precision dataset of building footprints to identify infrastructure deficits and infer informal settlements down to the street level everywhere in sub-Saharan Africa. We identify a general pattern of informality with urbanized areas showing, on average,  greater access to infrastructure and services than rural and periurban areas, each characterized by a statistically consistent spectrum of uneven local development.  We show that our physical measures of informality are systematically associated with many indicators of low human development, and that these form a single principal component predicted by specific functional changes of the built environment. These results demonstrate that the localization of sustainable development is possible down to the street level at a continental scale and provide a general distributed strategy for accelerating progress in infrastructure and service expansion that taps local innovations in a way that is equitable and context appropriate.
\end{abstract}
\begin{document}

\flushbottom
\maketitle
%
%
\thispagestyle{empty}

\noindent

The critical need for sustainable development has come into sharp focus recently, along with the ambition to address associated challenges all over the world in the decades ahead~\cite{keith_new_2022}.   These imperatives have lead to landmark international agreements to address climate change and achieve specific sustainable development goals such as the worldwide eradication of extreme poverty~\cite{mitlin_urban_2013}.   However, practical progress towards these goals, and the integrated science and technology necessary to support them, has remained slow. 
Two interlocking factors contribute to this situation. 
First, there is a mismatch of scales between commitments made by national governments and processes of human development, which take place primarily at smaller scales in human settlements and their local communities~\cite{mitlin_urban_2013,brelsford_heterogeneity_2017,porto_de_albuquerque_role_2021,elias_data_2022,sheth_measuring_2023}. Second, at these smaller geographic scales and especially in lower income nations, there is a lack of comparable, standardized, and sufficiently rich data~\cite{moreno_slums_2003,mitlin_urban_2013,  montgomery_urban_2008,blumenstock_fighting_2016,sahasranaman_life_2021}, leading to calls for innovations towards their generation and analysis~\cite{moreno_slums_2003,porto_de_albuquerque_role_2021,borel-saladin_data_2017}. 

Contrasting to these challenges, there has been significant recent progress in our scientific understanding of cities and neighborhoods~\cite{bettencourt_introduction_2021}. These environments not only house a growing majority of the world's population, but also naturally promote systemic societal change~\cite{brelsford_heterogeneity_2017,bettencourt_introduction_2021}. Specifically, the process of urbanization is associated with long run improvements in many interlocking facets of human development including higher real personal incomes, greater access to healthcare and education, and expanded public services~\cite{bettencourt_introduction_2021,brelsford_heterogeneity_2017,sahasranaman_life_2021}.  This is because of both urban network (agglomeration)  effects in socioeconomic activities, which increase their productivity, complexity and accelerate their results, and of economies of scale in infrastructure and service delivery characteristic of the denser built environment of cities \cite{bettencourt_growth_2007, bettencourt_origins_2013,bettencourt_introduction_2021}. Because of these general effects, it is easier, faster and more productive to extend socioeconomic opportunities, infrastructure, and services to populations in a larger city than it is to create them in the first place in small towns and rural areas. As a result, fast urbanization often induces increased rural to urban migration and results in a general pattern of infrastructure and service delivery spreading along the urban hierarchy, from larger cities to smaller settlements~\cite{brelsford_heterogeneity_2017,sahasranaman_life_2021}.  This dynamical pattern of development implies a sort of transition over space and time, where signatures of higher human development are nucleated unevenly with higher probability in better connected central locations in larger cities and spread from there, eventually to all their constituent local communities and other less urbanized regions. We will show below that this general pattern of development is characteristic also of sub-Saharan Africa, and a feature of its informal settlements ~\cite{brelsford_heterogeneity_2017,parnell_africas_2014}. 

As a part of this uneven dynamics of development, fast urbanization often becomes associated with the exacerbation of inequalities not only between larger cities and rural areas but also on smaller scales, between neighborhoods within each settlement~\cite{sampson_great_2012,brelsford_heterogeneity_2017,sheth_measuring_2023}. The most critical of these inequalities, because it entails many others, is the "challenge of slums" (or informal settlements)~\cite{moreno_slums_2003,parnell_africas_2014,mitlin_urban_2013,brelsford_toward_2018,porto_de_albuquerque_role_2021}. These are neighborhoods resulting from land settlement without coordinated urban infrastructure or legal frameworks~\cite{maseland_governance_2010}. As a result, informal settlements are almost always initially associated with multidimensional poverty, lack of basic services and insecure land tenure~\cite{moreno_slums_2003,mitlin_urban_2013,satterthwaite_building_2020}. In 2003, the United Nations declared slums "the face of 21st century urbanization", motivating the first global studies and the estimate of about 1 billion people living in informal settlements  worldwide. 
Central to addressing the problem was the enshrining of slum reduction and eventual eradication (by 2030) in the United Nations Millennium (goal 1) and Sustainable Development Goals (SDGs) target 11.1. Current assessments confirm that we are late to meet this goal and estimate that 1.1 billion people now live in informal settlements worldwide.

The present work started in support of data collections by local communities in self-declared slums~\cite{patel_knowledge_2012,patel_editorial_2012,mitlin_urban_2013,satterthwaite_building_2020}. In collaboration with federations of non-governmental and local communities, it developed a set of standardized surveys and detailed maps of buildings and public services in thousands neighborhoods in 18 countries and 224 cities~\cite{patel_editorial_2012,bettencourt_introduction_2021}.  As these and other methodologies --including remote sensing~\cite{jean_combining_2016,yeh_using_2020,prieto-curiel_scaling_2023,kohli_ontology_2012,friesen_similar_2018}, mobile phone traces~\cite{blumenstock_predicting_2015}  and crowd sourced mapping~\cite{soman_worldwide_2020}-- continued to improve, an essential spatial typology of informal settlements emerged  defined by buildings without street accesses. Lack of street access to informal places of residence and work is the proximate common cause of many physical and socioeconomic deficits, including the lack of official addresses and associated socioeconomic stigma, unavailability of emergency services (such as fire protection, ambulance) and disconnection from basic services, especially water and sanitation which are piped along public ways~\cite{brelsford_toward_2018, un-habitat_streets_2014,mitlin_urban_2013,satterthwaite_building_2020,beard_out_2022}. Thanks to extensive studies of informal settlements,  we now understand that this physical mismatch between buildings and street networks violates the basic principles of  urban built environments~\cite{brelsford_toward_2018,bettencourt_introduction_2021}. This violation of urban physical connectivity, entails a distinct cost-benefit relationship for slum residents, trading off the possibility of settlement in the present against lack of access to the network effects of cities. The result is an untenable situation of latent but temporarily stunted human development. Because this situation leads to high social costs and stresses not only for residents but also for their societies, it must be resolved, typically by the extension of urban infrastructure and services and the legal inclusion of such neighborhoods into the physical and socioeconomic fabric of cities. 

An additional difficulty deals with interpreting the function of urban built environments in terms of spatial data on buildings and streets networks. Over the last few years, topological methods capturing the detailed relational nature of buildings to streets --regardless of specific geometry-- have been proposed, tested empirically and implemented computationally. These advances now allow us, for the first time, to identify and characterize each street block systematically over vast regions of the world~\cite{brelsford_optimal_2017,brelsford_toward_2018,soman_worldwide_2020,bettencourt_introduction_2021}. 
Here we use the first complete dataset of building footprints for sub-Saharan Africa to take this analysis of informality and development to a continental scale. Our data consists of 5.4 million blocks, containing over 415 million buildings in 89 nations and 1404 cities (Table S1), characterizing the living environments of about 1.152 billion people across economically, culturally and geographically diverse settings ~\cite{parnell_africas_2014,maseland_governance_2010}. Along with the analysis of street access to buildings, we also characterize each block in terms of its number of buildings, their sizes and spatial densities and estimate its resident population by downscaling worldwide population raster maps to local street block geometries~\cite{montgomery_urban_2008}. This procedure creates a standardized, internationally comparable dataset supporting the localization of sustainable development metrics and population at the block level filling a substantial empirical gap, especially in Africa \cite{parnell_africas_2014,maseland_governance_2010,borel-saladin_data_2017}. We use this evidence to estimate informal settlement populations locally across sub-Saharan Africa and to quantify general patterns of human development connecting urban infrastructure networks to social services and human capabilities, including measures of health, education and income. 

\section*{Results}
We now show how high-precision maps of buildings and streets can be analyzed across scales to quantify the varying character of infrastructure deficits, informal settlements and associated interlocking facets of socioeconomic development. This analysis allows the simultaneous characterization of local contextual factors down to the street level and the statistics of informality across larger scales, from neighborhoods to cities, nations and the sub-continent. Particularly important will be differences by levels of urbanization, from urban cores to rural areas.   

\subsection*{Measuring infrastructure access across scales}
The definition of an informal settlement (slum) developed by the United Nations is "a settlement in which the majority of households experience one or more of the following deprivations: lack of secure tenure, lack of access to improved water sources, lack of improved sanitation facilities, insufficient living space, poor structural durability of the dwelling".  Except at the extremes, these properties are difficult to measure both in terms of access to the relevant information and because of inherent ambiguities leading to difficulties of classification of a neighborhoods as slum versus a non-slum~\cite{parnell_africas_2014,daniere_poverty_1999, mitlin_urban_2013}. A similar problem affects a growing number of studies attempting a binary classification of neighborhoods using machine learning applied to satellite and aerial imagery~\cite{leonita_machine_2018}. Some surveys evade these difficulties by relying instead on community self-identification and assessing their experience of existing services and living conditions~\cite{gulyani_living_2012,patel_knowledge_2012,mitlin_urban_2013,beard_out_2022,brelsford_heterogeneity_2017}. 

Inspired by work co-producing mapping and socioeconomic data about informal settlements with local organizations, we have created a simple but non-trivial criterion for identifying informality, which can be measured objectively and has deep roots in urban science: the lack of street access to buildings. This quantity is at once of practical and fundamental interest because it speaks to the processes by which cities assemble as self-consistent physical and social networks~\cite{bettencourt_origins_2013,bettencourt_introduction_2021}.  Water, electricity, drainage and sanitation are delivered to each building via adjacent streets together with emergency services. Addresses are also codified along municipal streets, regularizing many forms of socioeconomic recognition and access including rights and obligations of land tenure. For all these reasons, street access is not only a necessary condition for development but also a strategic policy solution with many co-benefits~\cite{un-habitat_streets_2014}. 

\begin{figure}[h!]
\centering
\includegraphics[width=\linewidth]{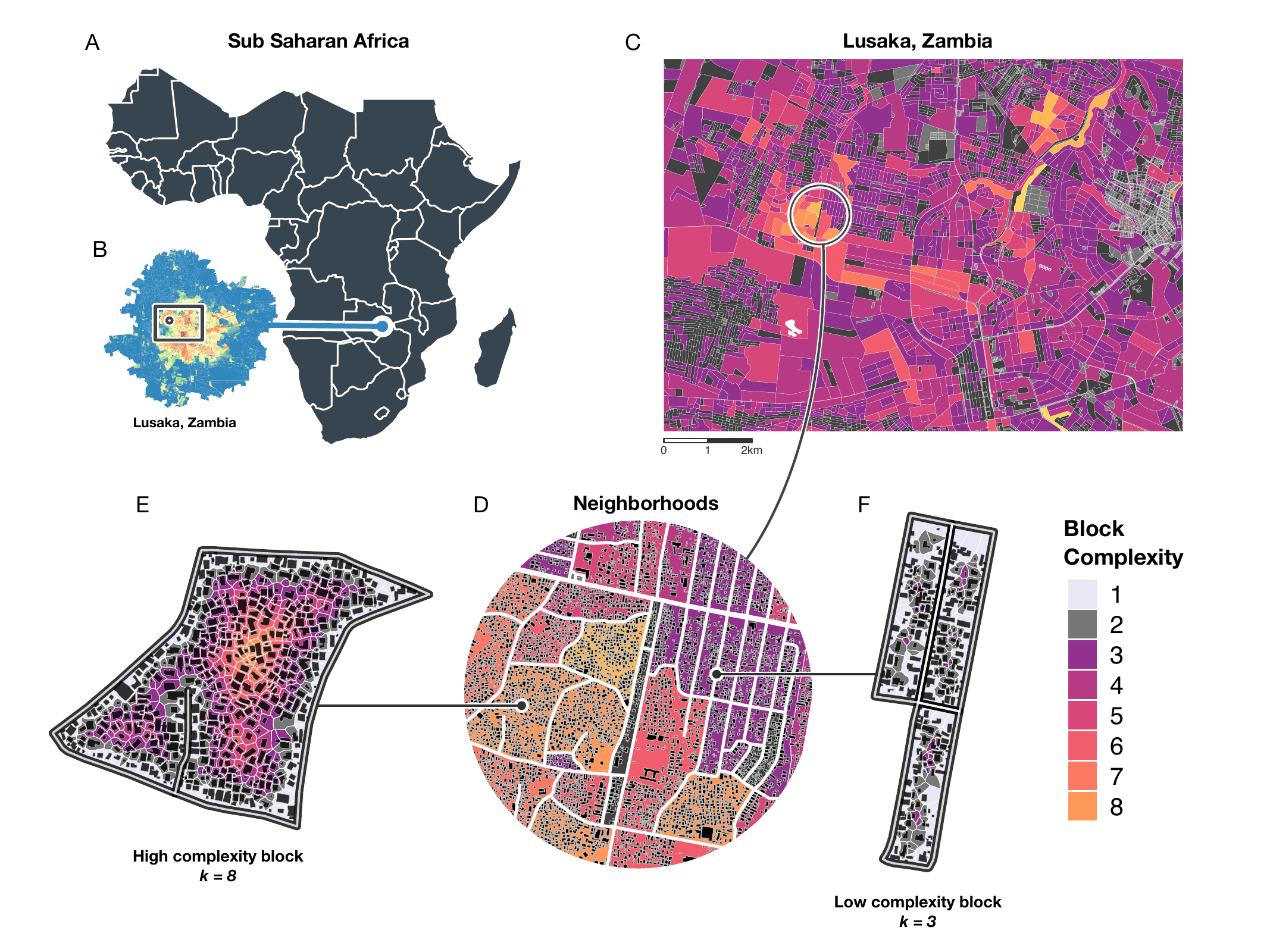}
\caption{{\bf The block decomposition of sub-Saharan Africa.} A. The decomposition of the subcontinent into nations, cities and rural areas. B. Lusaka (Zambia), the rectangular highlight is shown in Figure \ref{fig:1}C, and a more detailed regions showing all constituent blocks, in  D. We observe a large variety block types and shapes, delimited by streets shown in white. Building footprints are black polygons visible in the panels of Figure \ref{fig:1}E and F. For each block, a land parcel is identified around each building (thin white lines): these parcels form a graph expressing their spatial adjacency relationships, Figure~\ref{fig:s1}. Analysis of this block graph reveals how far each building is from the street network, denoted in different colors. The block complexity $k$ is the number of layers away from the street network for the most inaccessible building and denotes the difficulty of extending infrastructure and services. The block in  Figure \ref{fig:1}E is classified as informal by this measure, while the one in F is formal. See \url{https://www.millionneighborhoods.africa} for an interactive map.
}
\label{fig:1}
\end{figure}

Empirically, the (lack of) physical connection to buildings is a local feature of street networks, characterizing 'last mile' access. Because it is a relational quantity, it characterizes the spatial topology of cities regardless of their structural geometry~\cite{brelsford_toward_2018}. For example, it is independent of whether street plans are curvy or form a grid, and of the scale of blocks or buildings. These features make the identification of street access to buildings a mathematical problem in topology~\cite{brelsford_toward_2018,bettencourt_introduction_2021}. The relevant relation is the (unobstructed) adjacency of buildings to street networks and to each other, which can be captured via the construction of a block graph, Figures~\ref{fig:1} and \ref{fig:s1}.  The block graph is a network consisting of buildings (or land parcels) as nodes and their adjacency as edges. Network analysis of these block graphs identifies the access level of every building as the shortest path to any node at the street boundary shown by different colors in Figures \ref{fig:1} and \ref{fig:s1}. This approach is very efficient computationally because it decomposes the complex (and seemingly intractable) geography of nations and cities down to many independent blocks, which can be analyzed in parallel.

Recent work developed the mathematical and algorithmic methods to perform this analysis in general block geometries~\cite{brelsford_optimal_2017,brelsford_toward_2018} but remained limited to small scales by both data quality and computational efficiency. Here, we extend and improve these methods and apply them to new (complete) datasets of building footprints, street networks and population for the whole of sub-Sahararan Africa, characterizing every block across the subcontinent, see Methods. Table~\ref{tab:summary statistics} provides summary statistics.  

These methods quantify the severity of street access deficits to buildings in terms of a single topological number with a simple intuitive meaning. The block graph defines the summary statistic block complexity, $k$. This is the distance from the most internal building in the block to the nearest street access, measured as the minimum number of land parcels to be crossed. Block complexity values of $k=1,2$ denote universally accessible (or "planned") city blocks, where every building has direct street access or can do so via an easement between buildings. Values in the range $k=3-4$ are less accessible but can be the result of longer driveways, and the existence of non-residential backyard buildings. Numbers progressively higher characterize increasingly severe infrastructure deficits associated almost always with informal settlements as supported by systematic visual inspection, see Methods. Some false positives --blocks with large $k$ that are not informal settlements-- are sometimes found for particular types of institutional campuses, such as colleges, hospitals, or airports but these are very rare and easily identified by place names, building shapes and sizes. 

 Figure~\ref{fig:1} illustrates the general analytical procedure.  Using building footprints and street network data along with population raster maps (at larger scales, see Methods), we identify each city block as a geo-referenced polygon delimited by streets and other natural or human-made boundaries. This procedure is general and can be applied worldwide to produce uniquely identifiable block-level spatial units similar to the systems used to collect and report geographic census data in the U.S., or {\it setores censit\'arios} in Brazil.  Figure \ref{fig:1}A shows the decomposition of sub-Saharan Africa into nations and human settlements (cities and towns), with the city of Lusaka (Zambia) highlighted and expanded as an inset, Figure ~\ref{fig:1}B. The square box is then expanded in Figure \ref{fig:1}C, to reveal its decomposition in terms of blocks with colors revealing their $k$ complexity for a community area known as George. We observe a number of blocks identified as informal settlements by high block complexity (orange and yellow, left), as is also known from community surveys.  Other city blocks nearby are formal, in the sense that their buildings have direct street access, and correspondingly low $k$. Figure 1D shows details of informal and formal blocks, respectively. Informal settlement blocks are often larger spatially than nearby accessible blocks, and show clear lack of street access to internal buildings (black polygons). Table ~\ref{tab:summary statistics} and Figure~\ref{fig:s2} show that, on average, buildings in blocks with greater complexity are smaller, and that this is actually more prevalent in periurban and rural areas. Dead-end streets are common in informal settlements (Figure~\ref{fig:1}D, E) as an incipient means to create accesses to internal buildings at minimal cost~\cite{brelsford_optimal_2017,brelsford_toward_2018}, a process known as reblocking~\cite{brelsford_optimal_2017}. 

 In addition to measuring each block's complexity, we also produce block level estimates of ambient (resident and working) population. To do this, we projected population estimates from two different raster datasets in wide use - LandScan~\cite{rose_landscan_nodate} and WorldPop~\cite{linard_population_2012} - down to the block level using building footprint area, see Methods. In this way, we can characterize variations in local population size and density as well as building area (table \ref{fig:s1}), assessing crowding as estimated population per building area. The final ingredient of our analysis deals with the aggregation of blocks into human settlements and their characterization into settlement type categories: urban, secondary urban, periurban and rural. The classification of a block as urban denotes its inclusion in one city or town defined by the Global Human Settlements Layer (GHSL). Because these boundaries are most often constructed based on population density thresholds and spatial morphological features, they are likely to underestimate actual urban expansion at low densities in adjacent areas. To investigate this issue, we characterize blocks in the peripheries of urban areas by creating a periurban land classification, which comprises of a 10km buffer beyond the GHSL boundaries, corresponding to an easily commutable region to the central city. Secondary urban areas are smaller cities in the GHSL, which become associated to larger cities within this buffer and together form candidate regions for larger conurbations (Figures~\ref{fig:s4}).  Regions outside these three urban types, are classified as non-urban, or rural.

\subsection*{The statistics of informal settlements}
\begin{figure}[ht]
\centering
\includegraphics[width=\linewidth]{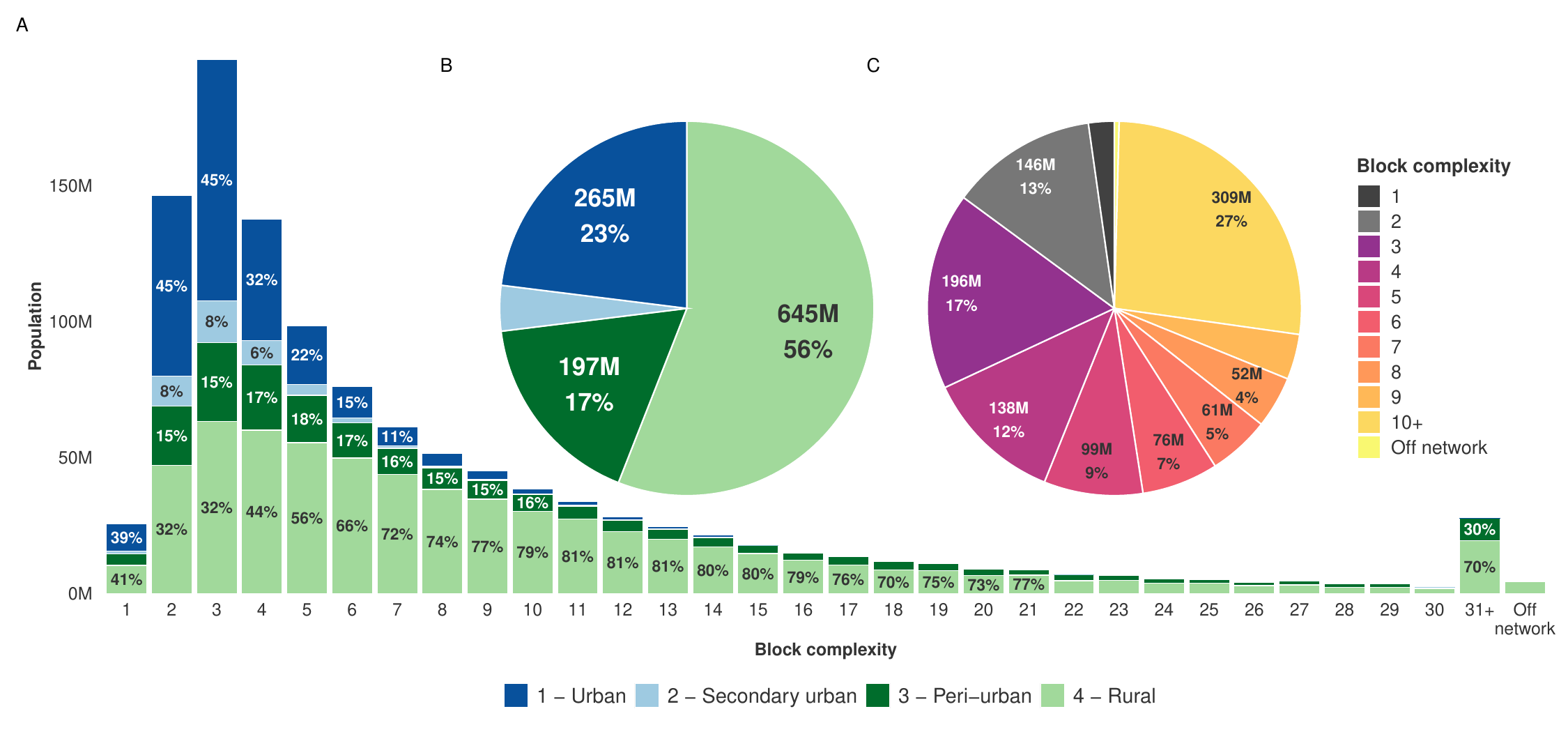}
\caption{{\bf The distribution of population in blocks with different complexity $k$, by urban levels across sub-Saharan Africa.} A. Histogram of total population in blocks with different $k$ and levels of urbanization. We see that extreme infrastructure deficits are relatively rare in cities and occur almost exclusively in urban peripheries and rural areas. B. The total population of sub-Saharan Africa by urbanization type, showing that the majority remains rural and periurban. C. Total population by levels of infrastructure deficits $k$, showing that about 50\% still live in blocks with strong deficits.}
\label{fig:2}
\end{figure}

Having characterized each block in each human settlement and rural region, we can now create and characterize the statistics of infrastructure deficits and inferred residential informality across scales.  Figure~\ref{fig:2} shows the frequency distribution of population living in blocks with different $k$ complexity, by settlement type and population size. The most salient feature of this distribution is that it is very broad, Figure \ref{fig:2}A. Figure \ref{fig:2}B shows the population decomposition in terms of type of settlement, indicating that by our estimate, the majority of the population in sub-Saharan Africa (56\%) remains rural. Importantly, transitional periurban areas, likely accounting for fast urban expansion, contain a total population around 200 million people, comparable to urban cores~\cite{mcgregor_peri-urban_2012}. Figure~\ref{fig:2}C shows the distribution of population totals across different levels of spatial access showing that more than half of the population of sub-Saharan Africa live in blocks with substantial or extreme lack of infrastructure accesses, a number in the range of 0.5-0.6 billion people. This is a new estimate for the total population living in informal settlements across sub-Saharan Africa but note that the majority is rural. 

Besides producing calculations of aggregate population totals, the distinct strength of these results lie in the localization of the challenge of informality at the block level, and providing an objective spatial measure of infrastructure deficits to estimate its severity. As such, our findings are somewhat more positive than an estimate of the total number of people living in informal settlements suggests. A fundamental takeaway from these results, made possible by the extensive character of the analysis, is that they do not support a simple dichotomy of neighborhoods classified as either slums versus non-slums. Instead, we observe a broad spectrum of access deprivation, with the most common neighborhoods in Figure \ref{fig:2}A (median $k=3$) actually having (almost) complete infrastructure access to each building. There is, however, also a long tail of neighborhoods whose buildings are much less accessible, including many extreme cases at $k>6$. Our results show that blocks with very high $k$ are actually quite rare in central urban areas, see also Figure~\ref{fig:s3}, though there are clearly some well known cases. This finding has the important consequence that many of the assumed characteristics of slums - being urban, high population densities, and crowding- are not typical of the general situation of informality in sub-Saharan Africa, which is much more rural and also periurban and at low density, Figure~\ref{fig:s2}. Surprisingly, crowding is much more likely in rural areas where most buildings are very small (median area 20$m^2$,  Figure~\ref{fig:s2}), despite very low population densities at the larger scales of blocks and regions, Table~\ref{tab:summary statistics}.

\begin{figure}[ht]
\centering
\includegraphics[width=\linewidth]{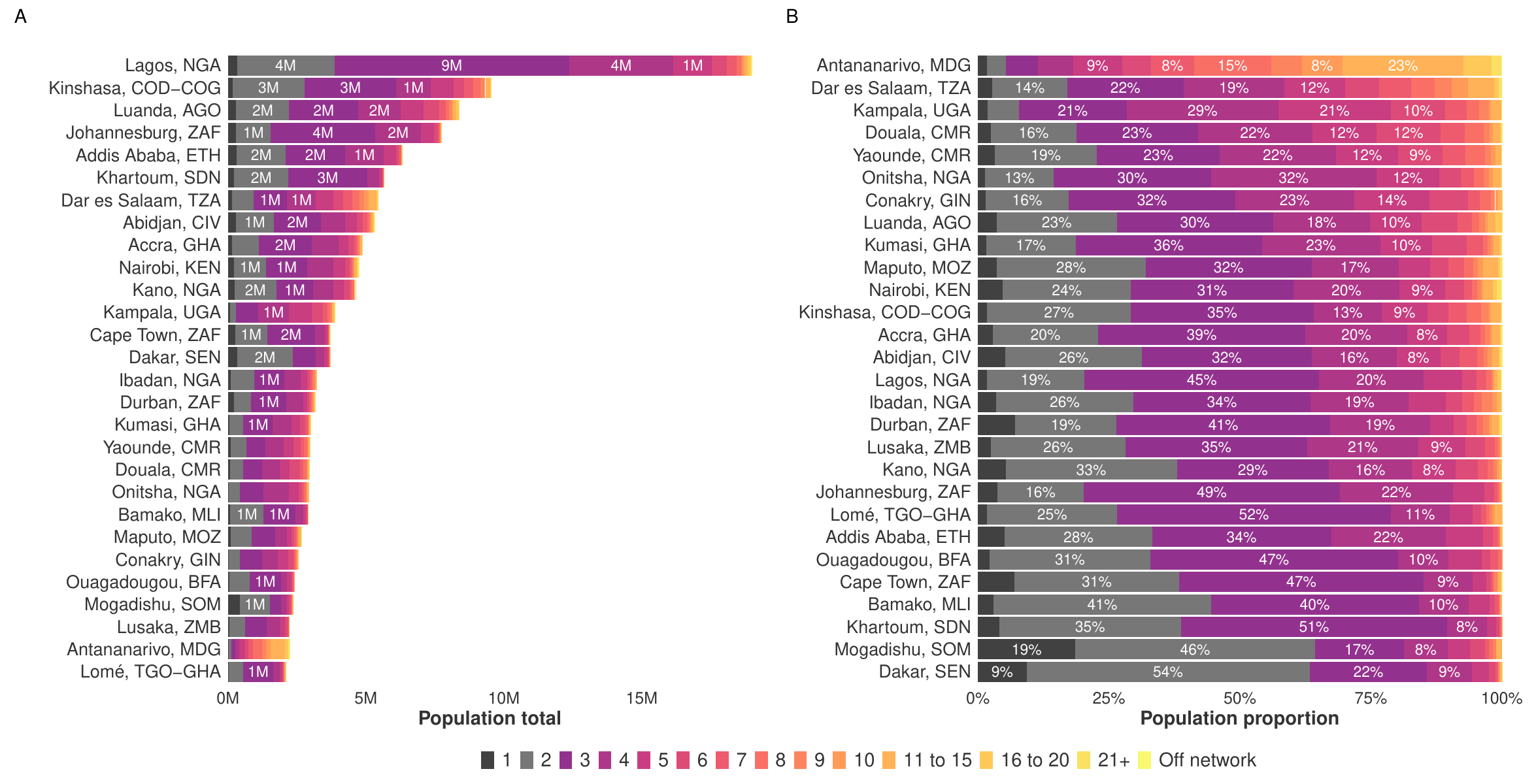}
\caption{{\bf Population spectrum of informality for the major cities of sub-Saharan Africa}. A. Cities ranked by population showing, for example, that Lagos (Nigeria) likely already provides infrastructure access to $>65$ \% of its population ($ k \leq 3 $). B. Cities ranked by higher levels of infrastructure deficits, $k$. Antananarivo (Madagascar) stands out as the subcontinents most informal city, but cities such as Dar Es Salaam (Tanzania) or Kampala (Uganda) also present significant challenges. Cities with the highest levels of access provision include Dakar (Senegal), Mogadishu (Somalia) and the cities of South Africa, but none in the continent has yet completely addressed the challenge of slums.}
\label{fig:3}
\end{figure}

Because blocks tile the entire territory, this type of analysis can be produced at any relevant aggregate scale including individual cities, regions, and nations. Figure~\ref{fig:3} shows the spectrum of block complexity, $k$, for each of the subcontinent's largest urban areas, see also Figure~\ref{fig:s4} for conurbations. The left panel shows urban areas sorted by population size with Lagos (Nigeria) being the largest in the subcontinent with about 22.5 million people. (Lagos is projected to grow to about 80 million by the end of the century~\cite{hoornweg_population_2017}.) The right panel shows cities sorted instead by higher levels of access deficits (higher $k$). We observe in general that larger African cities actually outperform their rural areas in infrastructure access, see also Figure \ref{fig:s3}-\ref{fig:s4}, but none has completely addressed the challenge of slums. Some cities such as Antananarivo (Madagascar) remain mostly informal, while others such as Dakar (Senegal), Cape Town or Johannesburg (South Africa) have provisioned more extensive street access, though many clearly identifiable spatial pockets remain (see interactive map, Methods).  The challenge is more severe for conurbations, because these enlarged city definitions include periurban areas where lower density informal settlements are common, Figure \ref{fig:s4}. Aggregating these trends at the national level (Figure \ref{fig:s5}) confirms that the least accessible blocks and the majority of population experiencing access deprivation is rural, but also identifies nations with substantially less access, such as Chad, Madagascar, Mozambique, Somalia, Sudan, and South Sudan.

\subsection*{Linkages between street access and human  development}
We now demonstrate that the block complexity $k$ has a deeper functional meaning, not only expressing spatial access deprivation but also entailing many other dimensions of low human development~\cite{parnell_africas_2014,mitlin_urban_2013}. This includes direct physical issues, such as lack of water and sanitation, but also a broad set of socioeconomic characteristics such as lower education, wealth and worse health.

At present there are no standard local data collections for human development indicators across sub-Saharan Africa~\cite{parnell_africas_2014,borel-saladin_data_2017}. Partially filling this gap, there is a long tradition of demographic and health surveys (DHS) supported by international agencies in collaboration with national statistics, see Methods. These surveys are less extensive than our block metrics, and moreover lack spatial precision. To compare the two types of evidence, we aggregated block data to subnational administrative subdivisions made available in the DHS survey data. The spatial scale of Figure \ref{fig:1}B gives a sense that this aggregation mixes together heterogeneous neighborhoods with significantly different characters~\cite{brelsford_heterogeneity_2017}, but we demonstrate below that correlations using these averages remain universally consistent and significant.

\begin{figure}[ht]
\centering
\includegraphics[width=\linewidth]{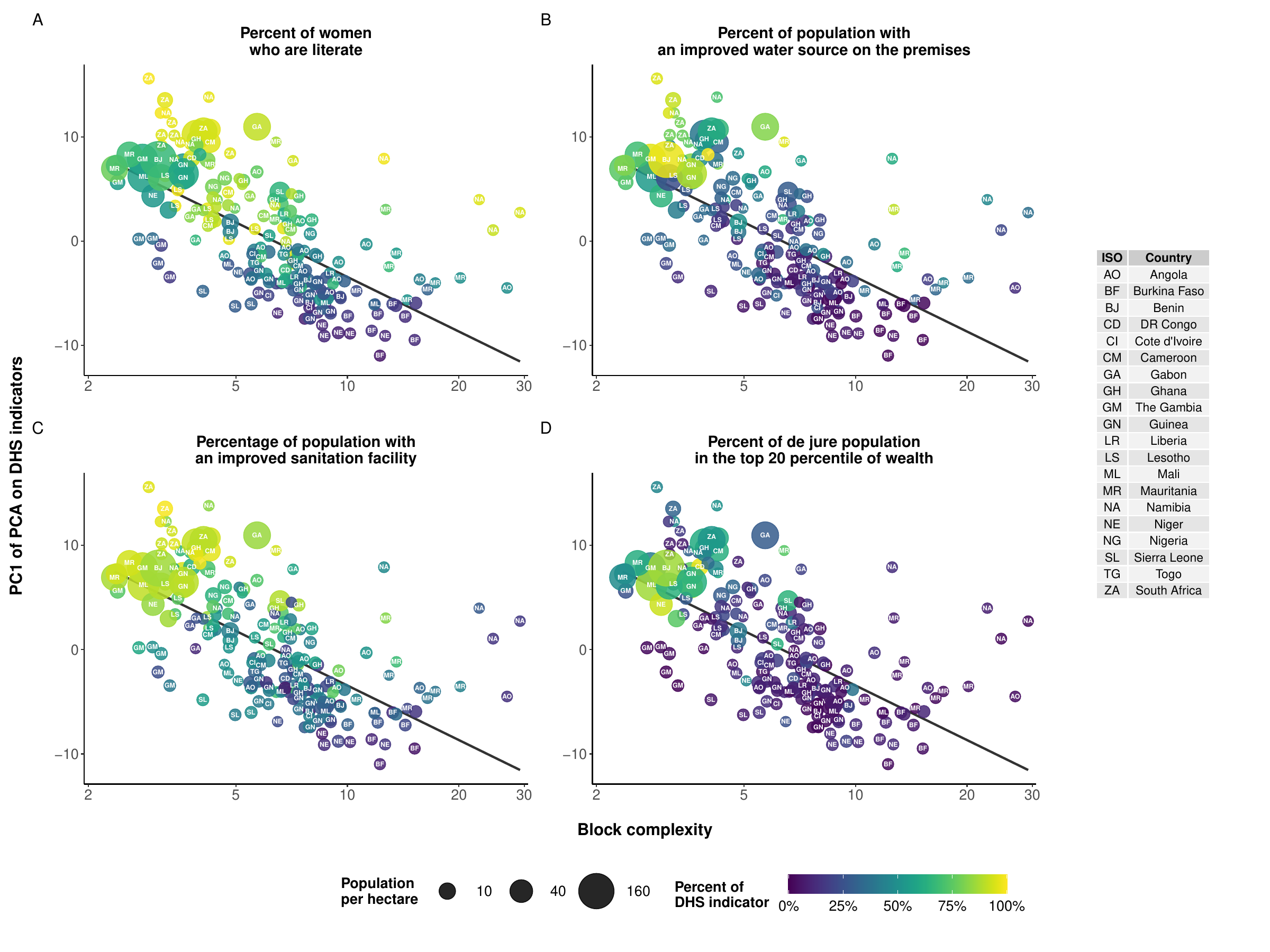}
\caption{{\bf The association between street access to buildings and different dimensions human development}. Measures of service provision, education and wealth are mutually correlated in demographic and health surveys and well characterized by the first component in a principal component analysis of 67 distinct variables, see Table \ref{tab:corrtable2}-\ref{tab:PCA}. Panels A-D show the variation of this component with block complexity $k$ at the regional level, across nations and urban areas. In all cases the relationship is negative showing that higher $k$ block complexity (and inferred informality) is systematically associated with lower female literacy, lower wealth and lower access to basic services, see also Figure~\ref{fig:s6}}
\label{fig:4}
\end{figure}

Using these spatially aggregated data, we performed a number of statistical analyses to establish the link between a large set of socioeconomic development measures and spatial access deprivation measured by block complexity, see Tables \ref{tab:corrtable2}-\ref{tab:PCA} and Figures \ref{fig:4} and \ref{fig:s6}. First, we correlated the variation of 67 different dimensions of human development on changes in block complexity, $k$, Table~\ref{tab:corrtable2}. We grouped these metrics in thematic groups including direct estimates of 
economic well-being, education and literacy, health, basic services, and household characteristics, including quality of housing and crowding. All these variables show systematic and significant correlations on block complexity across the subcontinent. For example, urban slum populations estimated at the national level are strongly correlated with higher average block complexity. Child mortality, underweight and stunting increase with larger $k$, while the fraction of live births in health facilities decreases. Higher block complexity is negatively correlated with measures of education at all levels (primary, secondary and higher) and median years of education. Higher spatial access deprivation is also associated with lower female literacy, see Fig. \ref{fig:4}A, though there are some notable exceptions such as Namibia, well known for its successful National Literacy Program, showing the potential for policy. Measures of access across all basic services, as expected, show a consistent pattern of increasing deficits with larger $k$, as do decreasing quality of housing measured by earth/sand and 'natural' floors. Measurements of crowding and wealth support this general picture associating multidimensional deprivations to block complexity, but the pattern of correlations adds interesting detail. For example, the fraction of households with 1-2 persons per sleeping room (no crowding) decreases with larger $k$, while larger numbers increase, as does the average number of persons per sleeping room. Parallel to these findings, the fraction of households in the two highest wealth quintiles decreases with $k$, while it increases in the lower 3 quintiles, especially the lowest. We also find that greater wealth inequality, measured by regional Gini coefficients, is associated with larger spatial access deprivation. Measures of economic well-being by consumption follow a similar pattern, with households with a refrigerator, private car, computer, television or mobile phone decreasing with $k$, the latter importantly showing a weaker negative correlation. 

All these statistical relationships are stable across nations and to the inclusion of control variables. In fact, these relationships become statistically much stronger (higher $R^2$) when treated at the national level via country fixed effects, and stronger still when we control for settlement type,  via the share of population living in urban areas, Table~\ref{tab:corrtable2}. 

Because we observe that these diverse measures of human development are mutually correlated, in agreement with studies in other local contexts~\cite{brelsford_heterogeneity_2017,sahasranaman_life_2021,sheth_measuring_2023}, we also characterized their joint variation with $k$. To do this, we performed a principal component analysis across regions and identified the leading collective dimensions of variation. The first principal component explains most of the variation ( $\sim 48$\%), Table \ref{tab:PCA}.  Figure~\ref{fig:s4} illustrates this behavior showing the variation of four different socioeconomic variables, specifically female literacy, access to water, sanitation and household wealth, see also Figure~\ref{fig:s6}. These results also show that denser (and explicitly more urban) regions perform better on average in all dimensions of human development relative to less dense and more rural regions. Table~\ref{tab:corrtable2}-\ref{tab:PCA} quantifies how country fixed effects and the share of urban population significantly improves these correlations, but that other factors such as building size, building to land area ratio, and population density matter relatively little  once block complexity and urban levels are factored in. 

We conclude that there is a clear and systematic pattern of multidimensional development across scales, with greater local street access and urbanization (at the regional level) facilitating socioeconomic opportunity, education, better health, and improved access to all basic services. The relatively larger multidimensional deprivations in rural and periurban areas speak to the necessity, as well as  the opportunity, to extend existing physical and socioeconomic access to these territories~\cite{maseland_governance_2010,mcgregor_peri-urban_2012}.  

\section*{Discussion}
We have shown that high-precision spatial data, specifically the combination of georeferenced building footprints and street networks, can be used to quantify physical access deficits, which in turn underlie an array of conditions associated with human development from basic services to heath and socioeconomic opportunity. We have shown that blocks with severe lack of street access to buildings, measured by the $k$ topological index, can be identified as informal settlements leading to a new objective method for producing localized street level estimates of population living in poverty, the main target of sustainable development goals 1.1 and 10.1 ~\cite{mitlin_urban_2013}.   

Where they have been mapped, this physical measure of lack of street access to buildings coincides with self-declared slums, and with multidimensional indicators of disadvantage measured by the best available international demographic and health surveys. While the present analysis deals with sub-Saharan Africa, we have created open-source tools that allow the generalization of these results to any other context where complete building footprints and street networks are available. It will be especially important in future work to extend the current analysis to other world regions, especially to Asia and South America where fast urbanization remains markedly informal but where comprehensive detailed characterizations are still lacking. It will also be critical to assess improvements in infrastructure delivery over time, to track localized sustainable development goals along with other local metrics of human development at the level of neighborhood communities. 

There are a few qualifications and caveats to the present results that we expect will be resolved as data and methodologies continue to improve. At present, building footprints are identified from manual tracing and remote sensing imagery using a combination of edge detecting algorithms, machine learning, and quality controls by human operators. This process achieves very good results, easy to verify against high precision images. Nevertheless, any analysis of this type fails to identify important information including building types, height, materials, and services all of which speak to issues of quality of life and human development~\cite{mitlin_urban_2013,brelsford_optimal_2017}.  Crowd-sourced data is also becoming excellent and may help fill these gaps but, at present, still remains incomplete and shows some inconsistencies especially at the smallest scales. Above all, these methods must become better integrated with the living experience of places known to local communities, civic organizations and local governments~\cite{patel_knowledge_2012,mitlin_urban_2013,satterthwaite_building_2020}. The convergence of these methods and perspectives is already gathering speed and scope towards producing richer, localized physical and socioeconomic data. Thus, we expect the kind of localized, systematic analysis introduced here to continue to improve and become standard in the very near future, providing a systematic framework for tracking and understanding deficits, inequalities, and successes in infrastructure and service delivery, development and urban evolution at the community scale everywhere on Earth. 

The systematic inference of informal settlements developed here points to a number of general features of urban development, which provide novel ingredients to innovation and policy. First, there is no basis for a classification of local communities as a dichotomy of "slums versus no slums", as has been assumed in both policy discussions and supervised classification methods from remote sensing imagery~\cite{satterthwaite_building_2020,kohli_ontology_2012,leonita_machine_2018}. Instead, our analysis points to the existence of a continuous spectrum of access deprivation, with most neighborhoods in sub-Saharan Africa experiencing a range of deficits of infrastructure and services, and only a relatively much smaller set being very deprived, denser and more complex, consistent with the findings of case studies~\cite{beard_out_2022}. Moreover, neighborhoods in the larger cities of sub-Saharan Africa tend to show fewer and less severe deficits of access infrastructure than in their corresponding rural and periurban areas, where, on the other hand, the nature of poverty and informality is very different~\cite{beard_out_2022,satterthwaite_building_2020}.  In this sense, the infrastructure deficits across sub-Saharan Africa present a number of diverse challenges that include relatively small but systemic improvements in most neighborhoods in urban areas, intense interventions in rarer but denser urban slums~\cite{gulyani_living_2012}, and comprehensive infrastructure development in relatively easier but extensive periurban and rural areas.  The block level maps introduced here transform these vast challenges into a much more treatable modular problem, where each block can be considered separately and whereby context appropriate solutions can be co-produced between resident communities and local governments. Other co-benefits, such as the generation of cadastral maps and addresses are naturally created by this process (Figure \ref{fig:1}C,D, \ref{fig:s1}) and can support context-appropriate institutions promoting secure land tenure, land use rights and responsibilities, and a formalized tax base for local governments~\cite{maseland_governance_2010}.

Despite a number of recent claims based on macroscopic indicators that African cities are lagging in terms of economic development, and that African urbanization is fundamentally different because of the abundance of informal settlements~\cite{kates_african_2007,pieterse_grasping_2011,parnell_africas_2014,turok_urbanization_2013,liddle_which_2015}, we would argue --without trivializing absolute levels of deprivation-- that the current evidence points in a different direction. We find that African cities, with variation as in Figure \ref{fig:3}, are doing significant better on average than their adjacent periurban and rural areas across every dimensions of development, including infrastructure delivery, basic services, healthcare access, and socioeconomic development~\cite{parnell_africas_2014}. Moreover, this gradient of development along the urban hierarchy, from larger to smaller cities and towns, is typical of past patterns of urbanization processes elsewhere~\cite{daniere_poverty_1999,bettencourt_introduction_2021,brelsford_heterogeneity_2017,sahasranaman_life_2021}.

Along with these practical considerations, the empirical evidence introduced here supports an emerging general understanding of the connection between urbanization and human development. Urban science has increasingly focused on the role of 
street networks supporting mobility and socioeconomic interactions, from which they derive socioeconomic value and predictable quantitative properties related to the population size of cities~\cite{bettencourt_origins_2013,bettencourt_introduction_2021,un-habitat_streets_2014}.  It is precisely the dividend of these connections 
that is only latent in fast developing cities with generalized infrastructure deficits.  As physical access networks become more present in the near future,  they are thus predicted to also expand and diversify socioeconomic networks and render cities more innovative and productive~\cite{bettencourt_introduction_2021}. Connected to this feedback between infrastructure and socioeconomic networks is an increasingly clear view of human development as the expansion of physical and socioeconomic connectivity. Thus, low human development is a personal condition of disconnection, associated with more local social networks dedicated primarily with coping and survival in the absence of supportive platforms, such as meaningful access to basic services, health care, public safety, and social institutions. This view of development as a partly physical network process, clearly more present in higher income cities, provides concrete strategies for human-centric policies that create virtuous cycles of change, whereby investments in (missing) physical connectivity can more than pay for themselves by generating broad socioeconomic development, which in turn can support stronger institutions, infrastructure and services~\cite{parnell_africas_2014}. 
    
The prospect of a general localized approach to sustainable development means that millions of neighborhoods around the world will be developing in parallel over the next decades, connecting to their local infrastructure networks and becoming increasingly formalized in the sense of public services, addresses and land uses, adding to their resilience to climate change and other stresses~~\cite{satterthwaite_building_2020}. This shared global experience brings online a vast peer to peer network of local innovators solving similar challenges, and supporting the creation of general knowledge and new technologies truer to the living experience of cities. In this way, we may leverage the growth of our scientific understanding of cities of all sizes and the growing possibilities of larger data to support and accelerate human development that is faster, more universal, and more sustainable. 

\section*{Methods}
\noindent {\bf Datasets:}
Building footprints data were produced by Ecopia Landbase Africa in 2022, powered by Maxar imagery and accessed via DigitizeAfrica Platform at \url{https://platform.ecopiatech.com}. Street network data were obtained from OpenStreetMap  available at \url{www.openstreetmap.org}, retrieved from the Geofabrik at \url{https://download.geofabrik.de/africa.html}. Population data were obtained from 2 worldwide raster maps by LandScan~\cite{rose_landscan_nodate}, available at \url{https://landscan.ornl.gov},  and WorldPop~\cite{linard_population_2012}, available at \url{https://data.worldpop.org/GIS/Population/Global\_2000\_2020\_Constrained}. Urban area geometries were based on the Global Human Settlements Layer  (GHSL) Urban Centre Database ~\cite{european_commission_joint_research_centre_description_2019} at \url{https://ghsl.jrc.ec.europa.eu}. Periurban areas are defined as a 10km spatial buffer around GHSL boundaries. National slum population estimates were obtained from the United Nations Human Settlement Programme (UN-Habitat) Global Urban Indicators Database 2020, available at \url{https://data.unhabitat.org/pages/housing-slums-and-informal-settlements}. Demographic and health survey data are available at \url{https://api.dhsprogram.com} and \url{https://spatialdata.dhsprogram.com}.

\noindent{\bf Block generation and population estimates:}
Block delineation is based on all connected streets in OpenStreetMap, from which we excluded the category of "footpaths" as these are reported irregularly and are not typically associated with infrastructure access. Blocks are defined as land geometries (polygons) circumscribed by street, roads and other boundaries such as rivers and coastlines. The code to create block geometries is available at \url{https://github.com/mansueto-institute/geopull}. Block level population estimates are produced by down-allocating population from spatial grids at larger scales (1km and 100m for LandScan and WorldPop, respectively) to each block proportionally to building area.\\
\noindent{\bf Validation, analysis and visualization:}
We inspected and validated building footprints against satellite imagery in a variety of locations to confirm (almost) full completeness and accuracy, except for very small buildings, which are likely not residential. We created an interactive visualization of the data and analysis that can also be used for local assessment at \url{www.millionneighborhoods.africa}, showing data for each block in terms of its $k$ complexity, estimated population, building counts and building and land areas. This also includes a population density map at high (block) precision, along with urban type classifications. Correlation analysis between these block characteristics and demographic and health indicators were obtained at subnational administrative units compatible with DHS surveys. Principal component analysis was performed over 67 indicators (dimensions) in 219 subnational region-year observations, as specified in Tables~\ref{tab:corrtable2} and \ref{tab:PCA}. The DHS data used in the analysis of correlations covers 238 administrative regions across 22 countries and 40 unique surveys taking place between 2010 and 2021, producing 367 unique survey observations. \\
\noindent{\bf Data and software:} The database is available online at \url{www.millionneighborhoods.africa/download}. Code for the analysis in this paper is available at \url{https://github.com/mansueto-institute/kblock-analysis}. Code for generating the underlying database is available at \url{https://github.com/mansueto-institute/kblock}.

\bibliography{MNM}

\section*{Acknowledgements}
We thank Io Blair-Freese and the Bill and Melinda Gates Foundation for access to the Ecopia building footprints data. We thank Anni Beukes,  Genie Birch, Christa Brelsford, and Pedro Concei\c{c}\~ao for discussions and audiences at the UN-Habitat Assembly and Slum Dwellers International for useful comments.  Dylan Halpern, Manuel Martinez, Merritt Smith, Idalina Sachango, Cooper Nederhood, and Satej Soman at the University of Chicago contributed to the underlying methods, code, and visualizations.

\section*{Author contributions statement}
LB, NM conceived the paper; NM performed data analysis and visualizations; LB, NM wrote the paper. All authors reviewed the manuscript. 

\newpage
\renewcommand{\thefigure}{S\arabic{figure}}
\setcounter{figure}{0}
\renewcommand\thetable{S\arabic{table}}    
\setcounter{table}{0} 

\section*{Supplementary Materials}

\subsection*{Supplementary Tables}
\begin{table}[ht]
\centering
\addtolength{\tabcolsep}{-0.2em}
\begin{tabular} {ccccccccc}
Urban &  Buildings & Population & Blocks  & Building  & Building  &  Building- &  Population & Block  \\ 
level &  number & (LandScan)  &  number &  area $m^2$ & density $ha^{-1}$ & block area ratio &  per building  & complexity \\
\hline
Urban & 60,863,684 & 265,447,669 & 1,506,939 & 97.12 & 16.67 & 0.16 & 4.36 & 3.70    \\
 {\small Secondary Urban} & 11,414,255 & 45,242,376 & 242,781& 106.78 & 13.04 & 0.139 & 3.96 & 3.68  \\
Periurban & 71,107,900 & 197,082,233 & 1,045,786& 63.65 & 0.58 & 0.0037 &  2.77 & 9.42  \\
Rural & 271,633,601 & 645,214,883 & 2,604,990 & 40.81 & 0.11 & 0.0004  & 2.38 & 9.65  \\
\hline
Total &   414,019,440 &  1,152,987,161  & 5,400,496  & 54.8 & 0.16 & 0.0009 & 2.78 & 8.01     \\
\hline
\end{tabular}
\caption{\label{tab:summary statistics} {\bf Summary Statistics of buildings, population and block statistics by urban level}. Note the changes in density from more to less urbanized areas and the opposite trend in service access expressed by rising block complexity. Note also the relative population weight and incipient infrastructure access in periurban areas.}
\end{table}


\begin{table}[htbp]
\centering
\renewcommand\theadalign{cc}
\renewcommand\theadfont{\normalsize}
\setcellgapes{2pt}
\makegapedcells
\begin{tabular}{|l|c|l|c|c|c|c|c|}
\hline
\thead{Group} & \thead{Subgroup} & \thead{Variable} & \thead{Spearman-\\rank\\Correlation} & \thead{N} & \thead{Used\\in\\PCA}  & \thead{PC1\\loading}   & \thead{PC2\\loading} \\
\hline \multirowcell{19}{\rotatebox[origin=c]{90}{Economic well-being}} 
& \multirow{8}{*}{\rotatebox[origin=c]{0}{Livelihoods}} 
& \% of men employed in agriculture & 0.6312 & 265 & Yes & -0.151 & -0.07 \\
& & \% of men employed in skilled manual labor & -0.3649 & 265 & Yes & 0.103 & -0.068 \\
& & \% of men employed in unskilled manual labor & -0.4669 & 245 &  &  &  \\
& & \% of men employed in prof., technical, mgmt. & -0.3811 & 265 & Yes & 0.088 & -0.003 \\
& & \% of women employed in agriculture & 0.4495 & 266 & Yes & -0.123 & -0.108 \\
& & \% of women employed in domestic labor & -0.4353 & 180 &  &  &  \\
& & \% of women employed in unskilled manual labor & -0.4308 & 246 &  &  &  \\
& & \% of women employed in prof., technical, mgmt. & -0.4632 & 266 & Yes & 0.118 & -0.074 \\

\cline{2-8} & \multirow{6}{*}{\rotatebox[origin=c]{0}{ \makecell{Wealth} }} 
& \% in the lowest wealth quintile & 0.4229 & 367 & Yes & -0.089 & -0.082 \\
& & \% in the second wealth quintile & 0.3702 & 367 & Yes & -0.086 & -0.157 \\
& & \% in the middle wealth quintile & 0.2655 & 367 & Yes & -0.037 & -0.119 \\
& & \% in the fourth wealth quintile & -0.221 & 367 & Yes & 0.072 & 0.077 \\
& & \% in the highest wealth quintile & -0.4888 & 367 & Yes & 0.105 & 0.168 \\
& & Wealth index Gini coefficient & 0.2956 & 363 & Yes & -0.112 & -0.091 \\

\cline{2-8} & \multirow{5}{*}{\rotatebox[origin=c]{0}{ \makecell{Household\\property} }} 
& \% of households possessing a computer & -0.6058 & 190 &  &  &  \\
& & \% of households possessing a refrigerator & -0.6721 & 367 & Yes & 0.156 & 0.062 \\
& & \% of households possessing a mobile telephone & -0.4358 & 367 & Yes & 0.12 & 0.092 \\
& & \% of households possessing a television & -0.7078 & 367 & Yes & 0.158 & 0.088 \\
& & \% of households possessing a private car & -0.6475 & 367 & Yes & 0.149 & 0.055 \\

\hline \multirowcell{22}{\rotatebox[origin=c]{90}{Education \& literacy}} 
& \multirow{8}{*}{\rotatebox[origin=c]{0}{Education}} 
& Median years of education (both sexes) & -0.5471 & 272 & Yes & 0.158 & -0.104 \\
& & Median years of education (males) & -0.5573 & 272 & Yes & 0.153 & -0.097 \\
& & Median years of education (females) & -0.5496 & 272 & Yes & 0.158 & -0.11 \\
& & \% of age 6+ with no education & 0.5329 & 272 & Yes & -0.138 & 0.16 \\
& & \% of age 6+ who attended secondary school & -0.6123 & 272 & Yes & 0.157 & -0.114 \\
& & \% of age 6+ who attended higher education & -0.6693 & 272 & Yes & 0.14 & 0.033 \\
& & Net primary school attendance rate & -0.5337 & 228 &  &  &  \\
& & Net secondary school attendance rate & -0.644 & 228 &  &  &  \\

\cline{2-8} & \multirow{14}{*}{\rotatebox[origin=c]{0}{ \makecell{Women\textquotesingle s\\education\\\& literacy} }} 
& Gross parity index for gross secondary school & -0.4939 & 228 &  &  &  \\
& & \% of women who are literate & -0.5743 & 367 & Yes & 0.148 & -0.135 \\
& & \% of young women who are literate & -0.5641 & 367 & Yes & 0.145 & -0.145 \\
& & \% of young women who cannot read at all & 0.5631 & 367 & Yes & -0.145 & 0.146 \\
& & Net primary school female attendance rate & -0.5415 & 228 &  &  &  \\
& & Net secondary school female attendance rate & -0.6931 & 228 &  &  &  \\
& & \% of women with secondary or higher education & -0.6283 & 367 & Yes & 0.161 & -0.128 \\
& & \% of females age 6+ with no education & 0.54 & 272 & Yes & -0.142 & 0.153 \\
& & \% of females age 6+ who attended secondary school & -0.6421 & 272 & Yes & 0.163 & -0.103 \\
& & \% of females age 6+ who attended higher education & -0.6444 & 272 & Yes & 0.143 & 0.004 \\
& & \% of women w/weekly access to newspaper, TV, radio & -0.4913 & 297 & Yes & 0.125 & -0.011 \\
& & \% of women with no access to mass media & 0.6695 & 297 & Yes & -0.141 & 0.006 \\
& & \% of women who have a bank account & -0.6586 & 96 &  &  &  \\
& & \% of women who own a mobile phone & -0.5492 & 114 &  &  &  \\

\hline
\end{tabular}
\label{tab:corrtable1}
\end{table}


\begin{table}[htbp]
\centering
\renewcommand\theadalign{cc}
\renewcommand\theadfont{\normalsize}
\setcellgapes{2pt}
\makegapedcells
\begin{tabular}{|l|c|l|c|c|c|c|c|}
\hline
\thead{Group} & \thead{Subgroup} & \thead{Variable} & \thead{Spearman-\\rank\\Correlation} & \thead{N} & \thead{Used\\in\\PCA}  & \thead{PC1}   & \thead{PC2} \\

\hline \multirowcell{13}{\rotatebox[origin=c]{90}{Health}} 
& \multirow{3}{*}{\rotatebox[origin=c]{0}{Women}}
& \% of women who are thin with $<$ 18.5 BMI & 0.4615 & 240 &  &  &  \\
& & \% of married women met need for family planning & -0.5551 & 266 & Yes & 0.121 & -0.156 \\
& & \% of live births delivered at a health facility & -0.5844 & 272 & Yes & 0.118 & -0.053 \\

\cline{2-8} & \multirow{10}{*}{\rotatebox[origin=c]{0}{Children}} 
& Child mortality rate & 0.4899 & 265 & Yes & -0.125 & -0.033 \\
& & Infant mortality rate & 0.378 & 266 & Yes & -0.096 & -0.08 \\
& & Postneonatal mortality rate & 0.4428 & 266 & Yes & -0.098 & -0.113 \\
& & Under-five mortality rate & 0.4614 & 265 & Yes & -0.121 & -0.061 \\
& & \% of 1 year olds who received all 8 basic vaccinations & -0.3309 & 271 & Yes & 0.039 & -0.093 \\
& & \% of children age 6-23 months fed 5+ food groups & -0.4641 & 266 & Yes & 0.112 & 0.025 \\
& & \% of children stunted ($<$-2 SD of height for age) & 0.5784 & 254 & Yes & -0.123 & -0.06 \\
& & \% of children underweight ($<$-2 SD of weight for age) & 0.6631 & 254 & Yes & -0.13 & 0.092 \\
& & \% of children wasted ($<$-2 SD of weight for height) & 0.4655 & 254 & Yes & -0.098 & 0.117 \\
& & \% of children under age 5 classified as having anemia & 0.4255 & 334 & Yes & -0.103 & 0.138 \\

\hline \multirowcell{15}{\rotatebox[origin=c]{90}{ Water, sanitation \& hygiene }} 
& \multirow{6}{*}{\rotatebox[origin=c]{0}{ \makecell{Sanitation\\\& hygiene} }} 
& \% of population with sanitation facility in dwelling & -0.447 & 127 &  &  &  \\
& & \% of population with basic sanitation service & -0.5484 & 367 & Yes & 0.125 & 0.061 \\
& & \% of population with an improved sanitation facility & -0.602 & 367 & Yes & 0.137 & 0.081 \\
& & \% of population using open defecation & 0.6534 & 367 & Yes & -0.098 & 0.006 \\
& & \% of population with soap \& water for handwashing & -0.4487 & 236 &  &  &  \\
& & \% of population with soap available for handwashing & -0.3566 & 236 &  &  &  \\

\cline{2-8} & \multirow{9}{*}{\rotatebox[origin=c]{0}{Water}} 
& \% of population with basic water service & -0.6031 & 367 & Yes & 0.134 & 0.1 \\
& & \% of population using improved water & -0.4811 & 367 & Yes & 0.122 & 0.107 \\
& & \% of population with improved water on premises & -0.6278 & 367 & Yes & 0.143 & 0.18 \\
& & \% of population with limited water service & 0.3798 & 367 & Yes & -0.062 & -0.006 \\
& & \% of population using an unimproved water source & 0.4818 & 367 & Yes & -0.122 & -0.106 \\
& & \% of population using water piped into dwelling & -0.4843 & 367 & Yes & 0.126 & 0.12 \\
& & \% of population with water under 30 min. round trip & 0.3133 & 367 & Yes & -0.113 & -0.205 \\
& & \% of population with water over 30 min. round trip & 0.4946 & 367 & Yes & -0.093 & -0.051 \\
& & \% of population with water on the premises & -0.5966 & 367 & Yes & 0.139 & 0.187 \\

\hline \multirowcell{15}{\rotatebox[origin=c]{90}{ Household Characteristics }} 
& \multirow{6}{*}{\rotatebox[origin=c]{0}{Crowding}} 
& Mean persons per sleeping room & 0.4202 & 367 & Yes & -0.079 & 0.292 \\
& & \% of households with 3 generations & -0.0503 & 367 & Yes & -0.018 & -0.015 \\
& & \% of households with 1-2 persons per sleeping room & -0.3612 & 367 & Yes & 0.069 & -0.293 \\
& & \% of households with 3-4 persons per sleeping room & 0.4115 & 367 & Yes & -0.091 & 0.246 \\
& & \% of households with 5-6 persons per sleeping room & 0.2102 & 367 & Yes & -0.042 & 0.273 \\
& & \% of households with 7 + persons per sleeping room & 0.1907 & 367 & Yes & -0.028 & 0.257 \\

\cline{2-8} & \multirow{5}{*}{\rotatebox[origin=c]{0}{Utilities}} 
& \% of population cooking in the house & -0.2343 & 266 & Yes & 0.113 & -0.043 \\
& & \% of population using clean fuel for cooking & -0.442 & 358 & Yes & 0.133 & 0.064 \\
& & \% of population using solid fuel for cooking & 0.4575 & 367 & Yes & -0.137 & -0.058 \\
& & \% of population with electricity & -0.6992 & 367 & Yes & 0.153 & 0.087 \\
& & \% of population with no electricity & 0.699 & 367 & Yes & -0.153 & -0.087 \\

\cline{2-8} & \multirow{2}{*}{\rotatebox[origin=c]{0}{Floors}} 
& \% of population with earth/sand floors & 0.5702 & 367 & Yes & -0.128 & -0.068 \\
& & \% of population with natural floors & 0.6058 & 367 & Yes & -0.134 & -0.056 \\

\cline{2-8} & \multirow{2}{*}{\rotatebox[origin=c]{0}{ \makecell{Slums} }} 
& \% of country's urban population living in slum & 0.6243 & 42 & & & \\ 
&& households, reported by UN-Habitat & & & & & \\ 
\hline
\end{tabular}
\caption{{
\bf Demographic and health surveys (DHS) subregional correlational and factor analyses on $k$ and other spatial variables.}}
\label{tab:corrtable2}
\end{table}


\begin{table}[htbp]
\centering
\renewcommand\theadalign{cc}
\renewcommand\theadfont{\normalsize}
\setcellgapes{2pt}
\makegapedcells

\begin{tabular}{lllllll}
\hline
\multicolumn{1}{l|}{\textbf{}} & \multicolumn{2}{l|}{\textbf{Univariate}} & \multicolumn{2}{l|}{\textbf{Country effects}} & \multicolumn{2}{l}{\textbf{ \makecell[l]{Country effects \&\\controls} }} \\ \hline
\multicolumn{1}{l|}{\textbf{Adjusted R$^2$}} & 0.2146 & \multicolumn{1}{l|}{} & 0.6098 & \multicolumn{1}{l|}{} & 0.8179 &  \\ \hline
\multicolumn{1}{l|}{\textbf{Covariate predictors of PC1}} & \textbf{Coefficient} & \multicolumn{1}{l|}{\textbf{P-value}} & \textbf{Coefficient} & \multicolumn{1}{l|}{\textbf{P-value}} & \textbf{Coefficient} & \textbf{P-value} \\ \hline
\multicolumn{1}{l|}{Intercept} & 4.7073 & \multicolumn{1}{l|}{0.0000} & 0.8503 & \multicolumn{1}{l|}{0.5036} & -4.4860 & 0.0003 \\
\multicolumn{1}{l|}{Block complexity} & -0.6590 & \multicolumn{1}{l|}{0.0000} & -0.6228 & \multicolumn{1}{l|}{0.0000} & -0.2996 & 0.0000 \\
\multicolumn{1}{l|}{Population share in urban areas} &  & \multicolumn{1}{l|}{} &  & \multicolumn{1}{l|}{} & 10.6840 & 0.0000 \\
\multicolumn{1}{l|}{Share of buildings under $31m^{2}$ } &  & \multicolumn{1}{l|}{} &  & \multicolumn{1}{l|}{} & 8.4182 & 0.5882 \\
\multicolumn{1}{l|}{Building footprint to block area ratio} &  & \multicolumn{1}{l|}{} &  & \multicolumn{1}{l|}{} & -36.1866 & 0.0449 \\
\multicolumn{1}{l|}{Population per hectare} &  & \multicolumn{1}{l|}{} &  & \multicolumn{1}{l|}{} & 0.0635 & 0.0550 \\
\multicolumn{1}{l|}{Cameroon} &  & \multicolumn{1}{l|}{} & 4.4184 & \multicolumn{1}{l|}{0.0007} & 4.2147 & 0.0000 \\
\multicolumn{1}{l|}{Guinea} &  & \multicolumn{1}{l|}{} & -0.1785 & \multicolumn{1}{l|}{0.8953} & -0.0409 & 0.9652 \\
\multicolumn{1}{l|}{Niger} &  & \multicolumn{1}{l|}{} & -2.0468 & \multicolumn{1}{l|}{0.2033} & -2.0022 & 0.0709 \\
\multicolumn{1}{l|}{Côte d'Ivoire} &  & \multicolumn{1}{l|}{} & 0.8060 & \multicolumn{1}{l|}{0.5870} & 1.2584 & 0.2196 \\
\multicolumn{1}{l|}{Gabon} &  & \multicolumn{1}{l|}{} & 7.0766 & \multicolumn{1}{l|}{0.0000} & 9.1024 & 0.0000 \\
\multicolumn{1}{l|}{Namibia} &  & \multicolumn{1}{l|}{} & 11.4452 & \multicolumn{1}{l|}{0.0000} & 12.4638 & 0.0000 \\
\multicolumn{1}{l|}{Sierra leone} &  & \multicolumn{1}{l|}{} & -0.1818 & \multicolumn{1}{l|}{0.9117} & 0.9420 & 0.4137 \\
\multicolumn{1}{l|}{Togo} &  & \multicolumn{1}{l|}{} & 1.3337 & \multicolumn{1}{l|}{0.4548} & 1.5926 & 0.1946 \\
\multicolumn{1}{l|}{DR Congo} &  & \multicolumn{1}{l|}{} & 0.9241 & \multicolumn{1}{l|}{0.5316} & 0.7654 & 0.4529 \\
\multicolumn{1}{l|}{Ghana} &  & \multicolumn{1}{l|}{} & 5.6949 & \multicolumn{1}{l|}{0.0002} & 5.7902 & 0.0000 \\
\multicolumn{1}{l|}{Lesotho} &  & \multicolumn{1}{l|}{} & 4.1297 & \multicolumn{1}{l|}{0.0084} & 6.9151 & 0.0000 \\
\multicolumn{1}{l|}{Angola} &  & \multicolumn{1}{l|}{} & 3.7609 & \multicolumn{1}{l|}{0.0041} & 2.6877 & 0.0032 \\
\multicolumn{1}{l|}{South Africa} &  & \multicolumn{1}{l|}{} & 12.5949 & \multicolumn{1}{l|}{0.0000} & 13.0673 & 0.0000 \\
\multicolumn{1}{l|}{Benin} &  & \multicolumn{1}{l|}{} & 1.6002 & \multicolumn{1}{l|}{0.2727} & 1.5403 & 0.1282 \\
\multicolumn{1}{l|}{Mali} &  & \multicolumn{1}{l|}{} & 0.4987 & \multicolumn{1}{l|}{0.7578} & 0.8045 & 0.4742 \\
\multicolumn{1}{l|}{Nigeria} &  & \multicolumn{1}{l|}{} & 4.9953 & \multicolumn{1}{l|}{0.0055} & 4.3431 & 0.0005 \\
\multicolumn{1}{l|}{The Gambia} &  & \multicolumn{1}{l|}{} & 3.0856 & \multicolumn{1}{l|}{0.0697} & 4.6341 & 0.0003 \\
\multicolumn{1}{l|}{Liberia} &  & \multicolumn{1}{l|}{} & 1.6457 & \multicolumn{1}{l|}{0.3811} & 2.8718 & 0.0296 \\
\multicolumn{1}{l|}{Mauritania} &  & \multicolumn{1}{l|}{} & 6.7953 & \multicolumn{1}{l|}{0.0000} & 6.6507 & 0.0000 \\ \hline

\multicolumn{7}{l}{Principal Component 1 (PC1) is the dependent variable and is taken from the PCA on 67 DHS indicators. PC1} \\
\multicolumn{7}{l}{explains 47.6\% of the variance in the PCA, followed by PC2 at 8.6\%. Regressions were based on 219 observations} \\
\multicolumn{7}{l}{ at the subnational administrative level. The Spearman-rank correlation between block complexity and PC1 is -0.709.} \\
\multicolumn{7}{l}{The model is a ordinary least squares regression.}
\end{tabular}
\caption{{\bf Principal component analysis of selected DHS variables versus block complexity $k$.} Note that the adjusted R$^2$ increases with country fixed effects and controls for other spatial variables. Of these only the region's level of urbanization (population share in urban areas) is significant.}
\label{tab:PCA}
\end{table}

\newpage 
\subsection*{Supplementary Figures}

\begin{figure}[ht]
\centering
\includegraphics[width=0.8\textwidth]{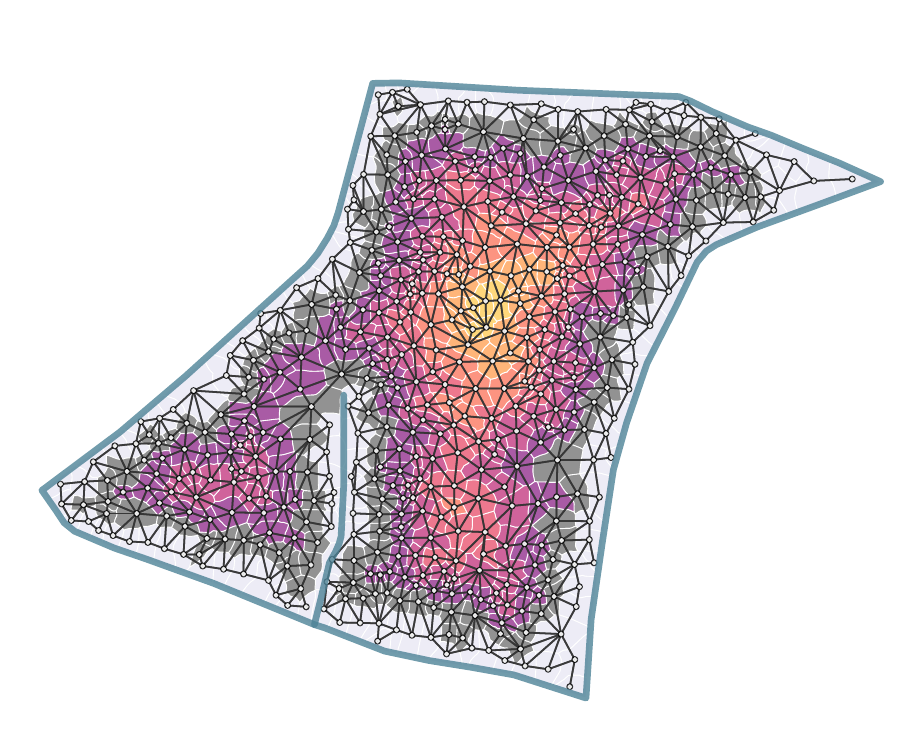}
\caption{{\bf Block graph for Figure~\ref{fig:1}E.} To create a block graph and calculate the block's $k$ complexity, we create a land parcel decomposition of the land area of the block, with each parcel encompassing a building (and its footprint), see methods. A the block graph is then created by connecting the center of each parcel to all spatially adjacent parcels. Parcels adjacent to streets are marked as external. Each parcel (building) is then characterized by its shortest distance to one of these external nodes in terms of number of links in the block graph, corresponding to parcels to be crossed. The block's $k$ complexity is longest such distance, corresponding to the shortest distance to the street network from the blocks least accessible parcel. As a matter of procedure, we create a parcel map in each block and a characterization of the distance to the street network of each building, shown in color gradient. The color characterizing each block in Figure \ref{fig:1}C, and in the interactive visualization map at \url{https://millionneighborhoods.africa} is that corresponding to this least accessible building. }
\label{fig:s1}
\end{figure}

\begin{figure}[ht]
\centering
\includegraphics[width=\linewidth]{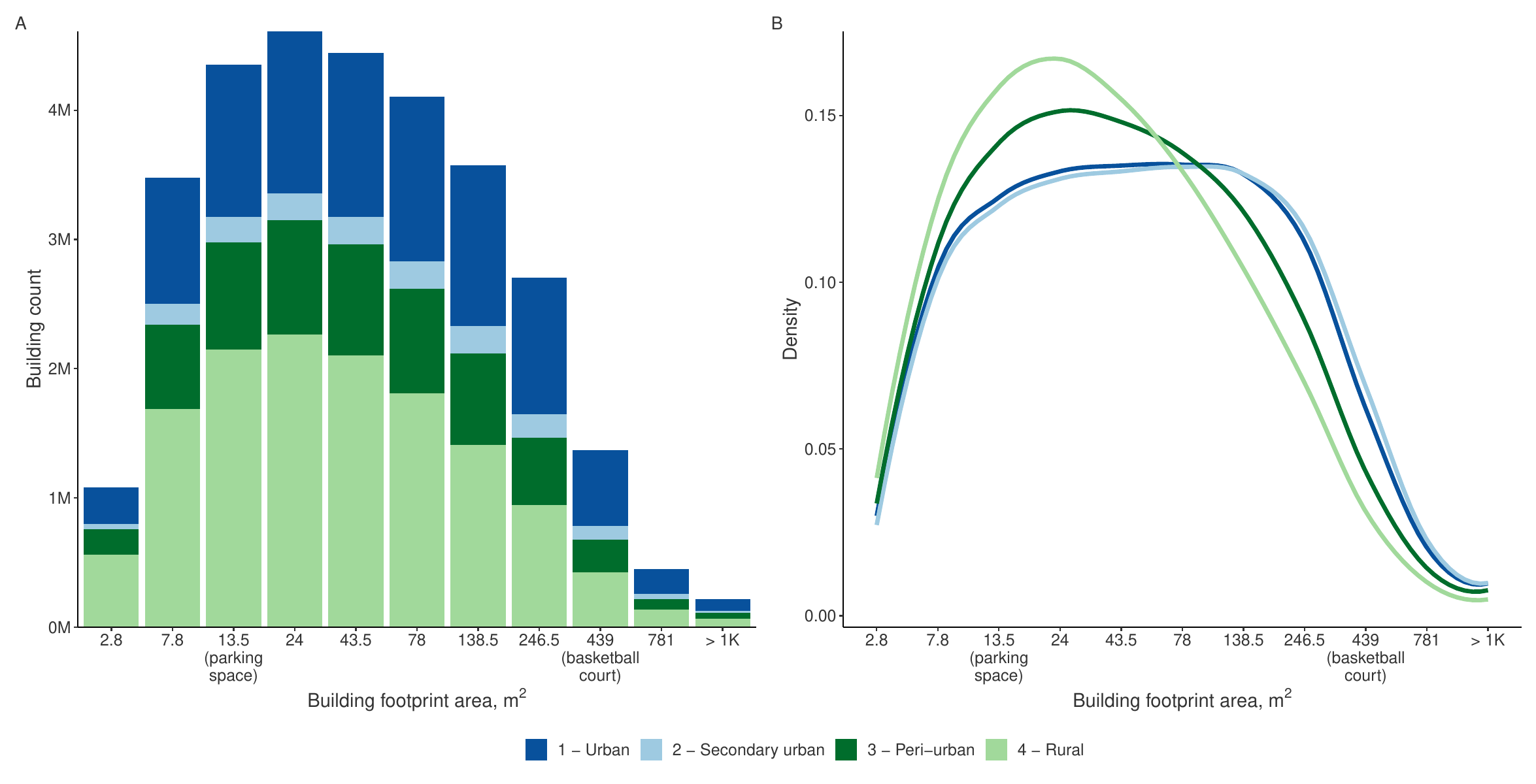}
\caption{{\bf The distribution of building footprint area, across sub-Saharan Africa by urban type}. A. Histogram of building numbers by area and urban type. B. The relative distribution of building sizes by urban type.  Note that rural areas are generally characterized by smaller buildings, with the median peaking around 20$m^2$. There is a much greater variety of building sizes across urban areas (principal and secondary) especially in the range 20-200$m^2$, where buildings are generally larger, see Table S1. Periurban areas have an intermediate character between urban cores and rural areas.}
\label{fig:s2}
\end{figure}

\begin{figure}[ht]
\centering
\includegraphics[width=\linewidth]{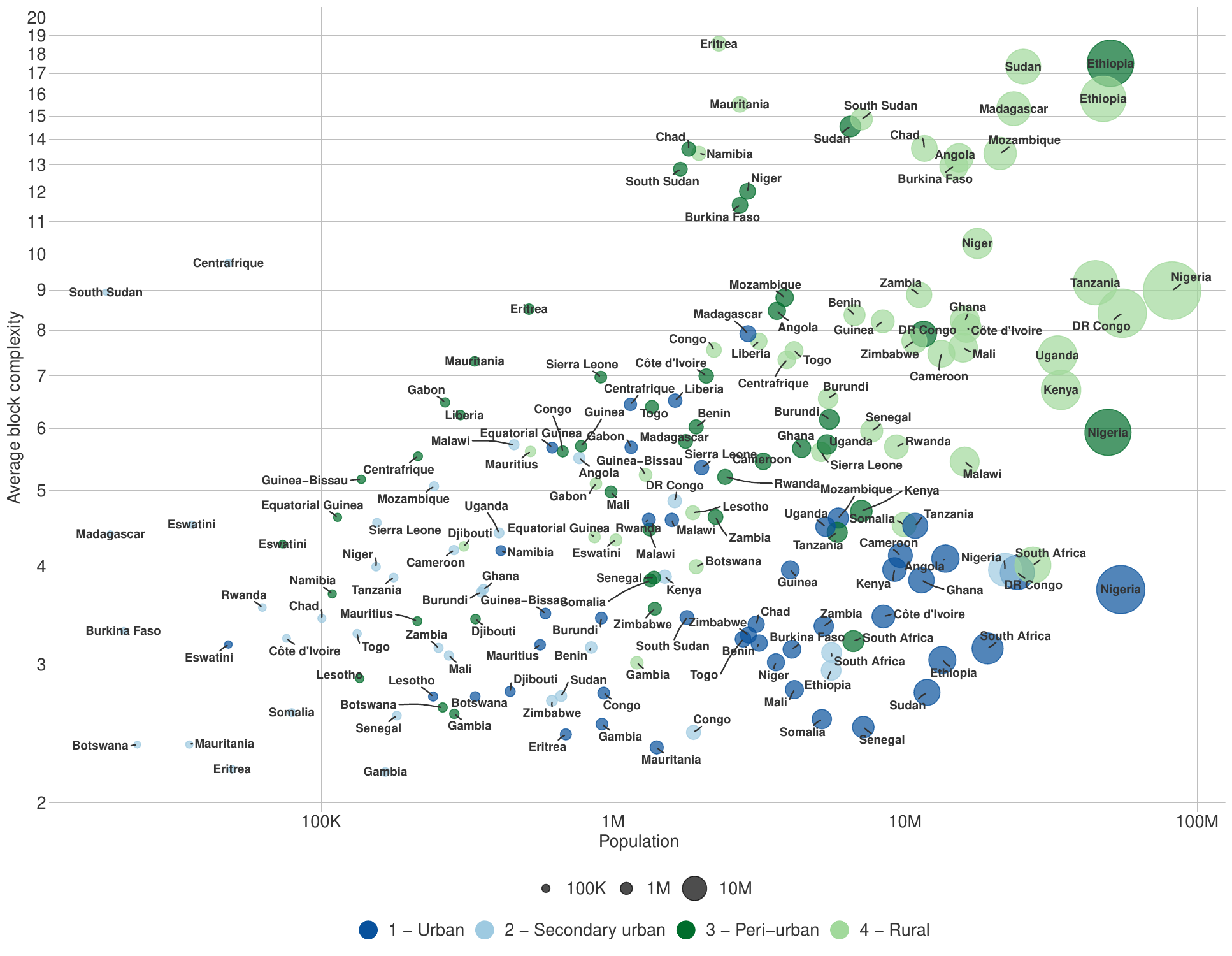}
\caption{{\bf The relationship between average block complexity and population size, by settlement type, urban to rural in nations of sub-Saharan Africa}. Note how lack of street access to buildings is predominantly a feature of rural areas, whereas urban areas in most nations have lower $k$, see also Table S1. Periurban areas show much larger variability, with good access in several nations (South Africa, Zimbabwe, Botswana), but manifestly low in many others (Sudan, South Sudan, Chad, Mozambique). These results suggest varying degrees of planning and infrastructure development across African nations, and also broader regions.}
\label{fig:s3}
\end{figure}

\begin{figure}[ht]
\centering
\includegraphics[width=\linewidth]{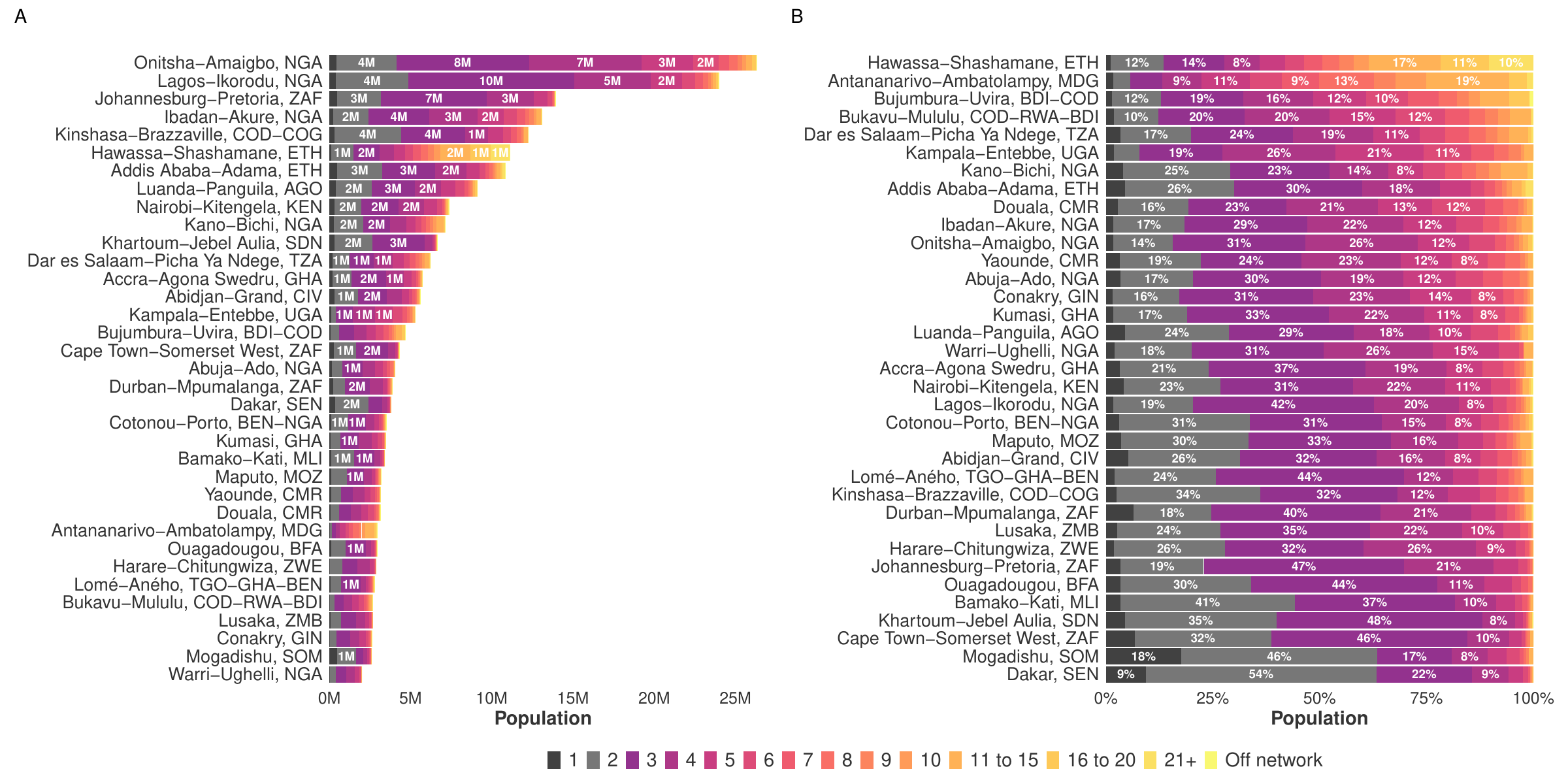}
\caption{{\bf The spectrum of informality across major sub-Saharan African conurbations}. This figure is analogous to Figure~\ref{fig:3}, but uses conurbations as its functional urban unit, instead of GHSL definitions. Conurbations group together GHSL cities with their 10km periurban buffer regions, and result naturally in larger urban areas. }
\label{fig:s4}
\end{figure}

\begin{figure}[ht]
\centering
\includegraphics[width=\linewidth]{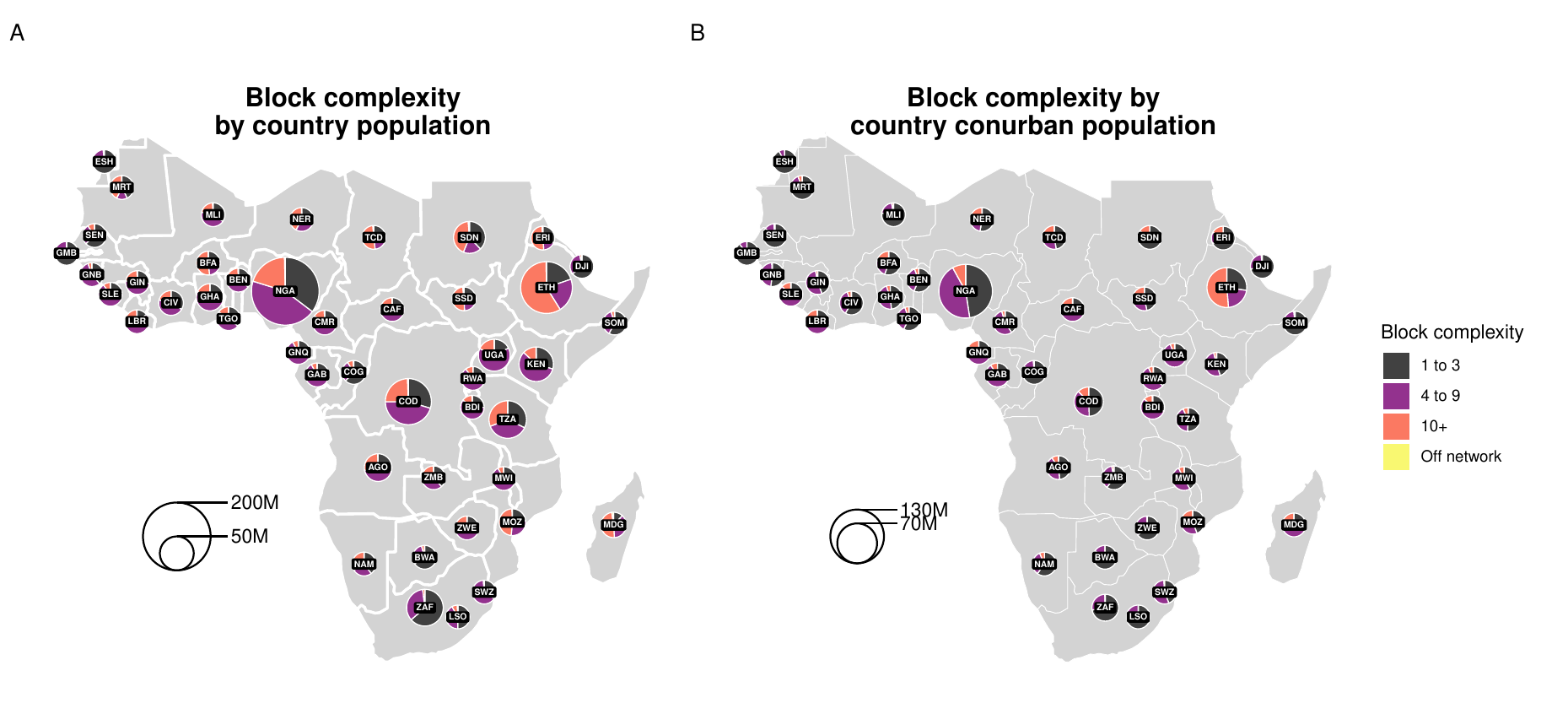}
\caption{{\bf Summary of distribution of block complexity across nation in sub-Saharan Africa}. The left panel shows results nationwide, while the left panel is limited to conurbations, see also Figure S3. We see significantly smaller block complexity in urban areas in general, and some nations with substantially less access than others, such as Madagascar, Mozambique, Somalia, Sudan, South Sudan and Chad.}
\label{fig:s5}
\end{figure}

\begin{figure}[ht]
\centering
\includegraphics[width=\linewidth]{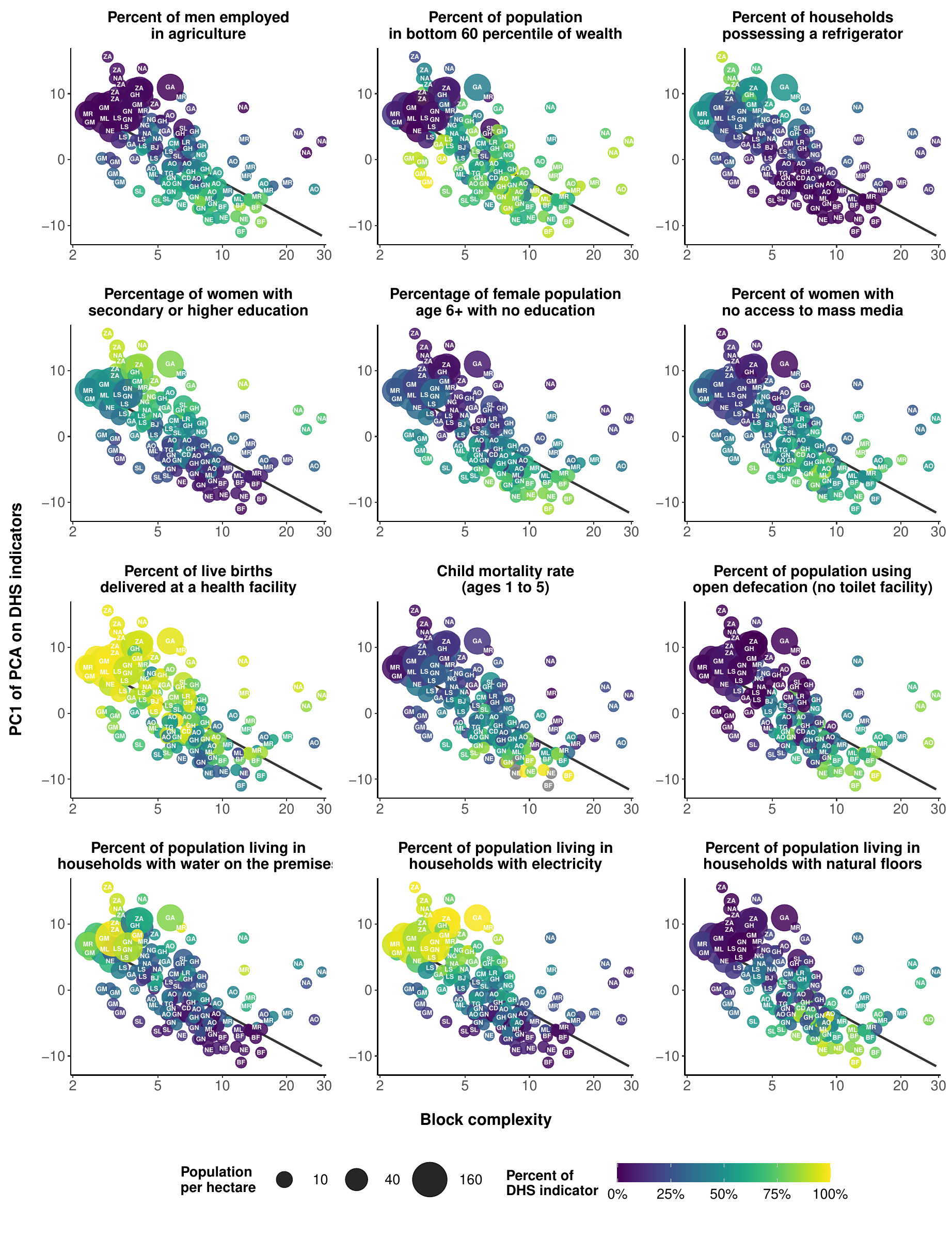}
\caption{{\bf The extended version of Figure~\ref{fig:4}, showing the correlation of a larger number of diverse indicators along their main principal component versus $k$ block complexity}. Symbol size denotes population density (denoting urban regions for larger symbols) and colors show the specific indicator variation along the PC1 variation with block complexity, $k$. We observe that measures of advantage (female literacy, access to services) are systematically anti-correlated with block complexity, while measures of disadvantage (poor housing quality, lower wealth) are correlated.}
\label{fig:s6}
\end{figure}

\end{document}